\newcommand{\eqn}[1]{\label{#1}}
\newcommand{\eq}[1]{\begin{equation}#1\end{equation}}
\newcommand{\eqnono}[1]{\begin{displaymath}#1\end{displaymath}}
\newcommand{\eqs}[1]{\begin{eqnarray}#1\end{eqnarray}}

\newcommand{\bra}[2]{{}_{#2}\big{<}#1\big{|}}
\newcommand{\ket}[2]{\big{|}#1\big{>}_{#2}}

\def\lb{\nonumber\\}

\newcommand{\binom}[2]{\left( \begin{array}{c} #1 \\ #2 \end{array}\right)}
\documentstyle[12pt]{article}

\def\d{\delta}

\def\l{\lambda}

\def\G{\Gamma}

\def\/{\over}
\def\*{\partial}
\def\|{\mid}

\def\hatpsi{\hat{\psi}}
\def\barS{\bar{S}}

\def\barM{\bar{M}}
\def\barpsi{\bar{\psi}}

\newcommand{\NP}[1]{Nucl. Phys.\ {\bf #1}\ }
\newcommand{\PL}[1]{Phys. Lett.\ {\bf #1}\ }

\newcommand{\NC}[1]{Nuovo Cimento\ {\bf #1}\ }

\newcommand{\IJMP}[1]{Int. J. Mod. Phys. \ {\bf #1} \ }
\newcommand{\half}{\mbox{\scriptsize $ \frac{1}{2}$}}
\newcommand{\third}{\mbox{\scriptsize $ \frac{1}{3}$}}
\newcommand{\tthird}{\mbox{\scriptsize $ \frac{2}{3}$}}

\begin{document}

\newtheorem{thm}{Theorem}
\newtheorem{prop}{Proposition}
\newtheorem{defn}{Definition}
\newtheorem{lemma}{Lemma}
\newpage
\pagenumbering{arabic}
\begin{flushleft}
G\"{o}teborg\\
ITP 93 - 3\\

February 1993
\end{flushleft}
\vspace{1.5cm}
\begin{center}
{\huge A Four-Reggeon Vertex for $ Z_3$ Twisted Fermionic Fields }\\[1cm]
{\large Niclas Engberg \footnote{TFENE@FY.CHALMERS.SE}\\
Bengt E.W Nilsson\footnote{TFEBN@FY.CHALMERS.SE}}\\[1cm]
{\sl Institute of Theoretical Physics\\Chalmers University of Technology\\
and University of G\"{o}teborg\\
S-412 96 G\"{o}teborg, Sweden}
\end{center}
\vspace{1cm}
\begin{abstract}
Using operator sewing techniques we construct the Reggeon vertex involving
four external
${\bf Z}_3$-twisted complex fermionic fields. Generalizing a procedure
recently applied to the
ordinary Ramond four-vertex, we deduce the closed form of the ${\bf Z}_3$
vertex
 by demanding it to reproduce the results obtained by sewing.
\end{abstract}
\newpage
Although a considerable amount of work has been devoted to using operator
sewing
techniques for string and conformal field theory (CFT) calculations
involving twisted fields, most of the attention has so far been focused on
${\bf Z}_2$ twisted fermions [1-16],
i.e. the Ramond sector of the NSR
string, and ${\bf Z}_2$ twisted scalars [17-19,6,20-25,12,26].
The applications of fields
with this basic twist (${\bf Z}_2$) are manifold, some of the more important
ones being besides string theory certain statistical models (see e.g.
\cite{ISZ}), and even some
of (so far) purely mathematical interest like the monster module
\cite{FLM1,CH2,DGM2}.
Even for this case of the simplest twist technical difficulties arise
already at the tree
level and for such a fundamental Reggeon vertex as the one with four
(external) twisted legs. For fermions these old problems
[2-6] were only
recently overcome \cite{ENS2,NS2}. In this short note we address the
generalization
of these questions to CFT's containing fields with higher twists
${\bf Z}_N,\;N>2$ [20-23].
It turns out that ${\bf Z}_N$ scalars play a very
important role when considering e.g. strings moving on orbifolds
\cite{HV1,DFMS}. However, generalizations of operator sewing methods to
CFT's with higher twist fields
(${\bf Z}_N,\;\;N>2$) have so far been (as far as we know) conspicuously
absent. We will
focus on ${\bf Z}_3$
twisted fermions for the simple reason that the corresponding case for
scalars is likely to possess additional complications which are not yet
understood even in the ${\bf Z}_2$ situation \cite{DGM2}. The goal here is
therefore to
demonstrate, using ${\bf Z}_3$ fermions, that the methods that were recently
successfully applied to the case of ${\bf Z}_2$ fermions in ref. \cite{ENS2},
and which led to that the
full four-Reggeon vertex can now be rigorously derived \cite{NS2}, can be
taken over to higher
twists. Although it turns out to be slightly more complicated to calculate
explicitly the quantities appearing in the ${\bf Z}_3$ twisted sewed vertex,
we are able to
deduce the closed form of the four-vertex. The answer generalizes the
previous results for the Ramond string \cite{DHMR,ENS2,NS2} in a very natural
way, and
further generalizations of our results to fermions with other twists is
probably straightforward.

The twisted four-vertex is constructed by sewing together two dual
four-vertices\footnote{See equation (\ref{dualvertex}) below.}
\cite{NTWH,NT1} emitting complex ${\bf Z}_3$ twisted
external fields with mode expansions
\eq{\hatpsi (z)=i\sum_{n \in {\bf Z}}\hatpsi_n z^{-n-\third }, \;\;\;\;\;\;
	\hat{\bar{\psi}}(z)=i\sum_{n\in {\bf Z}}\hat{\bar{\psi}}_n
	z^{-n-\tthird},\;\;\;\;\;\;
	\{\hatpsi_n ,\hat{\bar{\psi}}_m \} =\d_{n+m,0}}
and hermiticity properties
\eq{ \left( \hatpsi_n \right)^{\dagger}=\hat{\bar{\psi}}_n ,\;\;\;\;
 \left( \hat{\bar{\psi}}_n \right)^{\dagger}=\hatpsi_n \eqn{hprop} }
The ket vacuum is defined by
\eq{ \hatpsi_n \ket{0}{}=0, \;\;\;\;\forall n>0 \;\; ;\;\;\;\;
	\hat{\bar{\psi}}_n \ket{0}{}=0 ,\;\;\;\;\forall n\geq 0}
and the bra vacuum follows from (\ref{hprop}) and
$\bra{0}{}=(\ket{0}{})^{\dagger}$. The fermions may therefore be divided into
the following creation and annihilation parts
\eqs{\hatpsi^{(+)}(z)=i\sum_{n=1}^{\infty}\hatpsi_n z^{-n-\third}, \;\;\;\;
	\hatpsi^{(-)}(z)=i\sum_{n=0}^{\infty}\hatpsi_{-n}z^{n-\third} \\
	\hat{\bar{\psi}}^{(+)}(z)=i\sum_{n=0}^{\infty}\hat{\bar{\psi}}_n
	z^{-n-\tthird}, \;\;\;\;\hat{\bar{\psi}}^{(-)}(z)=i\sum_{n=1}^{\infty}
	\hat{\bar{\psi}}_{-n}z^{n-\tthird}}
with non-zero commutation relations $(|z|>|w|)$
\eq{ \{ \hatpsi^{(+)}(z),\hat{\bar{\psi}}^{(-)}(w)\}={-\left({w\/z}\right)^{
\third}\/
	z-w} \;,\;\;\;\;\;
	\{ \hat{\bar{\psi}}^{(+)}(z),\hatpsi^{(-)}(w)\}=
{-\left({z\/w}\right)^{\third}\/ z-w}}
The dual four-Reggeon vertex for complex fields is given by (see
\cite{NT1,ENS2,ENS1})\footnote{ From \cite{NT1} it
is clear that the derivation of the dual four-vertex is independent of the
twist
properties of the external field.}
\eq{\hat{W}(V)=\bra{0}{1}:e^{\oint dz \left[ \hatpsi_{aux}^{V}(z)
	(\hat{\bar{\psi}}(z)+i\hat{\bar{\psi}}_1 (z))+
	\hat{\bar{\psi}}_{aux}^{V}(z)(\hatpsi (z)+i\hatpsi_1 (z))\right]}:
	\ket{0}{1}\eqn{dualvertex}}
where $\hatpsi_1$ is the (untwisted) normal ordering field and the dots refer
only to the untwisted $aux$ field. As explained in \cite{ENS1} the normal
ordering
field must be untwisted to ensure that the vertex transforms correctly
under projective transformations (compare to the discussion in \cite{CO}
leading to the corresponding vertex in the NSR case). This means that when we
use the
Baker-Hausdorff (BH) formula to eliminate the normal ordering field, thereby
explicitly normal ordering the external fields
$\hatpsi (z),\hat{\barpsi}(z)$, the total
propagator in the BH term will not vanish. Instead this will give rise to a
term bilinear in the $aux$ field. This is the origin of the technical
problems that appear as soon as one tries to incorporate twisted fields.
The dual vertex describes the emission or
absorption from the world sheet of
two external twisted states located at $z_1 =V(\infty)$ and $z_2 =V(0)$. The
$aux$ field is taken to represent
the world sheet itself.

The four-vertex is constructed by multiplying two dual
vertices together and performing a vacuum correlation in the $aux$ field
\eq{\hat{W}_4 =\bra{0}{aux}\hat{W}_1 (V_1)\hat{W}_2 (V_2)\ket{0}{aux}}
By eliminating the normal ordering field and inserting the coherent state
completeness relation in the $aux$ Hilbert space the correlation becomes an
infinite
dimensional integral
\eq{\hat{W}_4=\int \prod_{r=\half}^{\infty}d^2 \psi_r
e^{-\half \tilde{\psi}^T G
	\tilde{\psi}+u^T \tilde{\psi}}}
where
\eqs{ \tilde{\psi}^T=\left( \begin{array}{cccc} \psi_r & \psi_{-r} &
\bar{\psi}_r &
	\bar{\psi}_{-r} \end{array} \right)\; ;\;\;\;
G=\left(\begin{array}{cccc}
 0 & 0 & M^{(++)}&-1\\ 0 & 0 & 1 & M^{(--)}\\-M^{(++)T} & -1 & 0 & 0 \\ 1 &
-M^{(--)T} & 0 & 0
\end{array}\right)}

\eq{u^T =\left(\begin{array}{cccc} \bar{U}_{r}^{(+)}(V_1) & \bar{U}_{r}^{(-)}
 ( V_2) & U_{r}^{(+)}(V_1) & U_{r}^{(-)}(V_2) \end{array}\right)}
Here the matrices $M^{(\pm \pm)}$ and vectors $U^{(\pm)}$,$\bar{U}^{(\pm)}$
are given by
\eqs{M^{(++)}_{rs}=-\oint_{0}dz \oint_{0} dw z^{-r-\half}
	{\left({V_{1}^{-1}(z)\/V_{1}^{-1}(w)}\right)^{\third}-1\/z-w}
	w^{-s-\half} \lb
	M^{(--)}_{rs}=-\oint_{\infty}dz\oint_{\infty}dw z^{r-\half}
	{\left(V_{2}^{-1}(z)\/V_{2}^{-1}(w)\right)^{\third}-1\/z-w}
	w^{s-\half}\lb
	\bar{U}_{r}^{(+)}(V_1)=\sum_{n\in {\bf Z}}\hat{\bar{\psi}}_n \oint_{0}
	dz\sqrt{V_{1}^{-1'}(z)}z^{-r-\half}(V_{1}^{-1}(z))^{-n-\tthird} \\
	\bar{U}_{r}^{(-)}(V_2)=\sum_{n\in {\bf Z}}\hat{\bar{\psi}}_n
	\oint_{\infty}dz\sqrt{V_{2}^{-1'}(z)}z^{r-\half}
	(V_{2}^{-1}(z))^{-n-\tthird}\lb
	U_{r}^{(+)}(V_1)=\sum_{n\in {\bf Z}}\hatpsi_n \oint_{0}dz
	\sqrt{V_{1}^{-1'}(z)}z^{-r-\half}(V_{1}^{-1}(z))^{-n-\third} \lb
	U_{r}^{(-)}(V_2)=\sum_{n\in {\bf Z}}\hatpsi_n \oint_{\infty}dz
	\sqrt{V_{2}^{-1'}(z)}z^{r-\half}(V_{2}^{-1}(z))^{-n-\third}\nonumber}
After performing the integral we find
\eq{\hat{W}_4 =det(G)e^{-\half u^T G^{-1}u}\eqn{Ires}}
Due to the projective invariance we can make a choice of our
projective transformations. This choice is arbitrary as long as the
transformations contain only one moduli parameter $\l$. The standard choice
is to let the emission points of the external
states be located at $\infty,-{1\/\l},-\l $ and $0$. This can be implemented
in four different ways \cite{ENS2} given in
Table \ref{Vchoices} where we have defined the matrix

\begin{table}
\eqnono{\begin{array}{|c|c|c|c|c|c|} \hline
 & z_{1}^{(1)},z_{2}^{(1)}& V_{1}^{-1} & M^{(++)} & \bar{U}_{rn}^{(+)}(V_1) &
	U_{rn}^{(+)}(V_1) \\
 & z_{1}^{(2)},z_{2}^{(2)}& V_{2}^{-1} & M^{(--)} & \bar{U}_{rn}^{(-)}(V_2) &
	U_{rn}^{(-)}(V_2) \\ \hline
I & -{1\/\l},\infty & z+{1\/\l} & -M^T &\bar{\psi}_{n}^{1}u_{r}^{(n)} &
	\psi_{n}^{1}v_{r}^{(n)} \\
& -\l ,0 & {1\/z}+{1\/\l} & M^T & -i\bar{\psi}_{n}^{2}u_{r}^{(n)} &
	-i\psi_{n}^{2}v_{r}^{(n)} \\ \hline
II & \infty ,-{1\/\l} & {1\/z+{1\/\l}} & M & i\bar{\psi}_{n}^{1}v_{r}^{-n} &
	i\psi_{n}^{1}u_{r}^{(-n)} \\
& -\l ,0 & {1\/z}+{1\/\l} & M^T & -i\bar{\psi}_{n}^{2}u_{r}^{(n)} &
	-i\psi_{n}^{2}v_{r}^{(n)} \\ \hline
III & \infty , -{1\/\l} & {1\/z+{1\/\l}} & M &
	i\bar{\psi}_{n}^{1}v_{r}^{(-n)} & i\psi_{n}^{1}u_{r}^{(-n)} \\
& 0,-\l &{1\/ {1\/z}+{1\/\l}} & -M & -\bar{\psi}_{n}^{2}v_{r}^{(-n)} &
	-\psi_{n}^{2}u_{r}^{(-n)} \\ \hline
IV & -{1\/\l},\infty & z+{1\/\l} & -M^T & \bar{\psi}_{n}^{1}u_{r}^{(n)} &
	\psi_{n}^{1}v_{r}^{(n)} \\
& 0,-\l & {1\/ {1\/z}+{1\/\l}} & -M & -\bar{\psi}_{n}^{2}v_{r}^{(-n)} &
	-\psi_{n}^{2}u_{r}^{(-n)}\\ \hline \end{array}}
\caption{Projective transformations with one modulus $\l$.\label{Vchoices}}
\end{table}
\eq{M_{rs}={r-{1\/6}\/r+s}\binom{-\third}{r-\half}\binom{-\tthird}{s-\half}
	\l^{r+s}\eqn{M}}
and the vectors
\eq{u_{r}^{(n)}=\l^{n+r+{1\/6}}\binom{-n-\tthird}{r-\half} \;,\;\;\;
	v_{r}^{(n)}=\l^{n+r-{1\/6}}\binom{-n-\third}{r-\half}}
For the choice denoted as $I$ we find that (\ref{Ires}) becomes
\eqs{\hat{W}_4 =det(1-M^2)
\exp{\left(\sum_{n,m \in {\bf Z}}
\left[ -\hat{\bar{\psi}}_{n}^1 u^{(n)T}M(1-M^2)^{-1}v^{(m)}\hatpsi_{m}^{1}+
i\hat{\bar{\psi}}_{n}^{1}u^{(n)T}(1-M^2)^{-1}v^{(m)}\hatpsi_{m}^2 -
\right. \right. }  \nonumber \\
\left. \left.
-i\hat{\bar{\psi}}_{n}^2 u^{(n)T}(1-M^2)^{-1}v^{(m)}\hatpsi_{m}^1-
\hat{\bar{\psi}}_{n}^2 u^{(n)T}M(1-M^2)^{-1}v^{(m)}\hatpsi_{m}^{2}\right]
\right) }
By using the matrix identity
\eq{A^{-1}(A-B)B^{-1}=B^{-1}-A^{-1} \eqn{TrivId}}
all quantities occurring in the exponent can be expressed in terms of the
quantities
\eq{ u^{T(n)}\xi^{(m)} \; ,\;\;\; u^{T(n)}\eta^{(m)} \;,\;\;\;n,m\in {\bf Z}}
where
\eq{ \xi^{(n)}=(1+M)^{-1}v^{(n)}, \;\;\;\; \eta^{(n)}=(1-M)^{-1}v^{(n)}
\eqn{defofxi}}
Similar quantities were introduced for the Ramond string in the 1970's
\cite{CGOS,SW2}
(for $n=m=0$) and generalized to arbitrary $m,n$ in \cite{ENS2}. Following a
procedure similar to that used for Ramond in reference \cite{CGOS} we will
now derive an
equation for $\xi =\xi^{(0)}$. For this purpose we introduce in addition
to $M$ in (\ref{M}) the
matrices (note that in the Ramond case only one type of M matrix is needed)
\eqs{\tilde{M}_{rs}={s-{1\/6}\/r+s}\binom{-\third}{r-\half}
	\binom{-\tthird}{s-\half}\l^{r+s}\\
	\barM_{rs}={r+{1\/6}\/r+s}\binom{-\third}{r-\half}
	\binom{-\tthird}{s-\half}\l^{r+s} \\
	\tilde{\bar{M}}_{rs}={s+{1\/6}\/r+s}\binom{-\third}{r-\half}
	\binom{-\tthird}{s-\half}\l^{r+s} }

and $R_{rs}=(r-{1\/6})\d_{r,s}$, $\bar{R}_{rs}=(r+{1\/6})\d_{r,s}$ such that
\eq{ R^{-1}MR=\tilde{M} \;\;\;\;\;\; \bar{R}^{-1}\barM\bar{R}=\tilde{\barM}}
Also
\eq{{\* M_{rs}\/\* \l}={Rvu^T \/\l}\;\;\;\;\;\;{\* \barM_{rs}\/\* \l}=
	{\bar{R}vu^T \/\l}}
and
\eq{M_{rs}+\tilde{\barM}_{rs}=vu^T \;\;\;\;\;\;\tilde{M}_{rs}+\barM_{rs}=vu^T}
We define vectors $\tilde{\xi}$,$\bar{\xi}$ etc. similarly as in
(\ref{defofxi}).
Using (\ref{TrivId}) with $A=1+\tilde{M}$ and $B=1-\barM$ gives
\eq{\tilde{\xi}={\bar{\eta}\/1+u^T \bar{\eta}}}
and with $A=1-\tilde{\barM}$, $B=1+M$ we find
\eq{\tilde{\bar{\eta}}={\xi \/1-u^T \xi}}
Then
\eq{{\* \xi \/\* \l}={R\tilde{\xi}\/\l}(1-u^T \xi)={R\bar{\eta}\/\l}
{1-u^T \xi\/1+u^T \bar{\eta}}}
Rearranging the factors, multiplying with $\l^{\third}$ and taking another
derivative with respect to $\l$
gives the equation
\eq{{\* \/\* \l}\left( \l^{4\/3}{1+u^T \bar{\eta}\/1-u^T \xi}
	{\* \xi \/\* \l}\right)={\* \/\* \l}
	\left( R\bar{\eta}\l^{\third}\right)={R\bar{R}\/ \l^{\tthird}}
	\tilde{\bar{\eta}}(1+u^{T}\bar{\eta})={R\bar{R}\/ \l^{\tthird}}
	\left({1+u^T \bar{\eta}\/1-u^T \xi}
	\right) \xi \eqn{Diffeq}}
To find $\xi_r$ we must make an ansatz for the unknown functions $u^T \xi$ and
$u^T \bar{\eta}$. By examining their Taylor expansions we find that
\eq{u^T \xi =1-{(1-\l )^{\tthird}\/(1+\l)^{\third}},\;\;\;\;
	u^T \bar{\eta}=-1+{(1+\l)^{\third}\/(1-\l)^{\tthird}}\eqn{ansatz}}
Inserting these functions in (\ref{Diffeq}) gives the solution for $\xi_r$
\eq{\xi_r ={\l^{r-{1\/6}}\/(1+\l)^{2r-\third}}\binom{-\third}{r-\half}{}_2 F_1
(r,r-{1\/6},2r+1,{4\l \/(1+\l)^2})\eqn{xir}}
The next logical step would be to use the solution (\ref{xir}) to calculate
$u^T\xi$ and $u^T\bar{\eta}$ thus verifying the ansatz (\ref{ansatz}).
Unfortunately, to perform the summation over $r$ is more complicated in this
case than for the Ramond string. To see why this is so let us recall what
happens for Ramond. For the Ramond string the solution for the
corresponding
quantity $\xi_{r}^R$ is \cite{CGOS,SW2}
\eq{\xi_{r}^R ={\l^{2r}\/(1+\l)^{2r}}\binom{-\half}{r-\half}{}_2 F_1
(r,r,2r+1,{4\l\/(1+\l)^2})}
By using the integral representation for the hypergeometric function
$\xi_{r}^R$ becomes, through a miraculous cancellation of $\G$-functions,
\eq{\xi_{r}^R ={1\/\sqrt{2}}\oint_{C}{dt\/2\pi it}\omega (\l,t)^r\;,\;\;\;\;\;
\omega (\l,t)={t-{(1+\l)^2 \/ 4\l}\/t(1-t)}\eqn{xiR}}
The function $\omega (\l,t)$ is such that it removes
the square root cuts in the sum leaving only poles. The simplification that
occurs in (\ref{xiR})
does not have a direct analogue in the ${\bf Z}_3$ case. For ${\bf Z}_3$ it
seems
like we would
need a function with the ability to remove cubic root cuts. This could
hopefully be achieved by some change of variables in the integral
representation of the hypergeometric function but this has so far eluded us.
However, we have checked that
the Taylor expansions agree to high order in $\l$ making an error unlikely.

The closed form of the Ramond vertex was first suggested in \cite{LeCl2} for
three-vertices and generalized to an arbitrary number of legs in \cite{DHMR}
using
path integral arguments. A direct sewing verification of these expressions was
given in \cite{ENS2} for the four-Ramond vertex where it was shown how
to calculate the quantities occurring in the exponent of the sewed vertex at
any oscillator level.
It was finally proved rigorously for the four-Ramond vertex in \cite{NS2}.
The complex four-Ramond vertex can be written\footnote{ Inside the
correlation functions the normal ordering is w.r.t. the bosonic field in
terms of which the spin fields $S,\bar{S}$ and fermion fields $\psi,\barpsi$
are bosonized. This normal ordering is thus completely independent of the one
indicated by the double dots outside the exponential and which refers to the
external Ramond fields $\psi_R ,\barpsi_R$. Note that the bars on the
external fields have moved compared to reference \cite{ENS2}.
This is because we have placed the zero modes differently.}
\eqs{\hat{W}_{R_1 ,R_2}=
\bra{0}{}:\barS(z_{2}^{(1)})S(z_{1}^{(1)})::\barS(z_{2}^{(2)})
S(z_{1}^{(2)}):\ket{0}{} \nonumber \\
:\exp{(\oint_{C_{z}}dz\oint_{C_{w}}dw\hat{\psi}^{V^{-1}}_{R}(z)
\bra{0}{}\psi(z)\barpsi(w):\barS(z_{2}^{(1)})S(z_{1}^{(1)})\barS(z_{2}^{(2)})
S(z_{1}^{(2)}):\ket{0}{}\hat{\barpsi}^{V^{-1}}_{R}(w))}:}
where $\psi$ and $S$ are the bosonized fermion and spin fields respectively.
The $z_{i}^{(j)}$ are the insertion points of the external states
corresponding to any of the choices listed in Table \ref{Vchoices}. The
$\hatpsi_{R}^{V^{-1}}$'s are sums of the external twisted fields transformed
with their respective projective transformations. The contours $C_z$ and $C_w$
enclose all branch cuts appearing in $\hat{\psi}_{R}^{V^{-1}}(z)$ and
$\hat{\barpsi}_{R}^{V^{-1}}(w)$ respectively. This expression of course
suggests a natural closed form of the complex ${\bf Z}_3$ twisted vertex. We
just replace the spin fields $S(z)=S_{\half}(z)$ by their ${\bf Z}_3$
analogues, the twist fields  $ S_{\third}(z)=:e^{{i\/3}\phi (z)}:$.
Thus we propose the following form of the ${\bf Z}_3$ four-Reggeon vertex:
\eqs{\hat{W}_4 =
\bra{0}{}:\barS_{\third}(z_{2}^{(1)})S_{\third}(z_{1}^{(1)})::
\barS_{\third}(z_{2}^{(2)})
S_{\third}(z_{1}^{(2)}):\ket{0}{} \nonumber \\
:\exp{(\oint_{C_{z}}dz\oint_{C_{w}}dw\hat{\psi}^{V^{-1}}(z)
\bra{0}{}\psi(z)\barpsi(w):\barS_{\third}(z_{2}^{(1)})S_{\third}(z_{1}^{(1)})
\barS_{\third}(z_{2}^{(2)})
S_{\third}(z_{1}^{(2)}):\ket{0}{}\hat{\barpsi}^{V^{-1}}(w))}:\eqn{ClosedformS}}

\begin{table}
\eqnono{\begin{array}{|c|c|c|}
\hline
n,m & I & IV \\ \hline
0,0 & i\l (1-\l^2)^{-\third} & -\l^{\tthird} \\ \hline
1,0 & i\l^2 (1-\l^2)^{-1} & -\l^{5\/3}(1-\l^2)^{-\third} \\ \hline
0,1 & i\l^2 (1-\l^2)^{-\third}(1+{\l^2 \/3}(1-\l^2)^{-1}) &
-\l^{-\third}(1-{\l^2 \/3}) \\ \hline
1,1 & i\l^3 (1-\l^2)^{-1}(1+{4\l^2 \/3}(1-\l^2)^{-1}) &
(1-\l^2)^{-\third}({4\l^{8\/3}\/3}-\l^{\tthird}) \\ \hline
-1,0 & i(1-\l^2)^{-\third}(1-{2\l^2 \/3}) &
-\l^{-\third}(1-{\l^2 \/3}) \\ \hline
0,-1 & i & -\l^{5\/3}(1-\l^2)^{-\third} \\ \hline
-1,-1 & i({1\/\l}+{\l\/3}) &
(1-\l^2)^{-\third}({4\l^{8\/3}\/3}-\l^{\tthird}) \\ \hline
-1,1 & i\l^2 (1-\l^2)^{-\third}\{ (1-{2\l^2 \/3})({1\/\l}+
{\l\/3}(1-\l^2)^{-1})-\l\} & -\l^{-{4\/3}}-{\l^{\tthird}\/3}-{\l^{8\/3}\/9} \\
\hline
1,-1 & i\l (1-\l^2)^{\third} & -\l (1-\l^2)^{-{5\/3}} \\ \hline
\end{array}}
\caption{Coefficients of $\hatpsi_{n}^1 \hat{\barpsi}_{m}^2$ for two
different choices of projective transformations.\label{closedres}}
\end{table}
We have verified that this is indeed the correct form of the vertex by
calculating the functions occurring in the exponent for level zero and one
oscillators (see Table \ref{closedres} for some examples) and comparing their
Taylor expansions
to those obtained directly
from the sewing expression (\ref{Ires}). The expansion of the determinant
prefactor in (\ref{Ires}) also agrees with the correlation in
(\ref{ClosedformS}) indicating that
\eqs{det(1+M^T M)=det(1+MM^T)=(1-\l^2)^{-{1\/9}} \\
	det(1-M^2)=(1-\l^2)^{1\/9}}
If we rewrite (\ref{ClosedformS}) in terms of the projective transformations
$V_{1}$ and $V_{2}$ we find the alternative closed form of the vertex
\eq{\hat{W}_4 =\left( {V_{2}^{-1}(z_{1}^{(1)})\/ V_{2}^{-1}(z_{2}^{(1)})}
\right)^{1\/9}
:e^{\oint_{C_z}dz\oint_{C_w}dw \hat{\barpsi}^{V^{-1}}(z)G(V_1 ,V_2 ,z,w)
\hat{\psi}^{V^{-1}}(w)}:\eqn{ClosedformV}}
where the propagator $G(V_1 ,V_2 ,z,w)$ is given by
\eq{G(V_1 ,V_2 ,z,w)={\left( {V_{1}^{-1}(z)\/ V_{1}^{-1}(w)}{V_{2}^{-1}(z)\/
V_{2}^{-1}(w)}\right)^{1\/ 3}\/ z-w}}
Although we have not proved that this is the closed form of the
vertex we feel that there could be little doubt as to its correctness.
A rigorous proof
along the lines of \cite{NS2} should be possible. The technical obstacles
encountered do not appear insurmountable and remain objects of further study.
It should also be
straightforward to perform the steps presented here for fermions with
arbitrary ${\bf Z}_N$ twists.

\hfill
\vspace{1cm}

${\bf Acknowledgement:}$

We are grateful to Per Sundell for many useful discussions.

\end{document}